\documentclass[aps,showpacs,12pt,superscriptaddress]{revtex4} 

\usepackage{amsmath}
\usepackage{graphicx}
\usepackage{subfigure}

\begin{document}

\title{Modulational Instability of Nonlinearly Interacting \\ Incoherent Sea States}

\author{P. K. Shukla} 
\altaffiliation[Also at: ]{Centre for Nonlinear Physics, Department of Physics, 
Ume{\aa} University, SE--901 87 Ume{\aa}, Sweden}
\altaffiliation{Centre for Fundamental Physics, Rutherford Appleton Laboratory,
Chilton, Didcot, Oxfordshire, UK}
\affiliation{Institut f\"ur Theoretische Physik IV and Centre for Plasma Science and Astrophysics,
Ruhr-Universit\"at Bochum, D-44780 Bochum, Germany}

\author{M. Marklund}
\email{mattias.marklund@physics.umu.se}
\altaffiliation[Also at: ]{Centre for Fundamental Physics, Rutherford Appleton Laboratory,
Chilton, Didcot, Oxfordshire, UK}
\affiliation{Centre for Nonlinear Physics, Department of Physics, 
Ume{\aa} University, SE--901 87 Ume{\aa}, Sweden}

\author{L. Stenflo}
\affiliation{Centre for Nonlinear Physics, Department of Physics, 
Ume{\aa} University, SE--901 87 Ume{\aa}, Sweden}

\date{Submitted to JETP Letters Oct.\ 3, 2006; 
  resubmitted Nov.\ 2, 2006; 
  accepted Nov.\ 17, 2006}

\begin{abstract}
The modulational instability of nonlinearly interacting spatially incoherent Stokes 
waves is analyzed. Starting from a pair of nonlinear Schr\"odinger equations, we derive 
a coupled set of wave-kinetic equations by using the Wigner transform technique. 
It is shown that the partial coherence of the interacting waves induces 
novel effects on the dynamics of crossing sea states. 
\end{abstract}

\maketitle

Extremely large amplitude freak waves (also referred to as rogue or giant waves) in oceans 
are well-known \cite{r1,r2,Kharim-Pelinovsky,r3}. Freak waves, which are steep waves 
and which can appear quickly from a relatively calm sea, are responsible for the loss of many 
human lives and ships. Therefore, it is important to understand the nonlinear 
propagation \cite{r4,r5,r6,r7,r8} of such waves in dispersive fluids that are far from a 
stationary state.  Specifically, we recall that in non-stationary oceans, the dynamics of 
modulated Stokes waves is under certain circumstances governed by a nonlinear Schr\"odinger 
(NLS) equation \cite{r10}, from which so called freak waves can occur. We note that 
the occurrence of freak waves may also be due to, e.g.\ a linear superposition of harmonic 
waves or wave-current interactions, and can thus be described by a different set of equations, 
such as the Zakharov equation \cite{Fedele}. The latter depicts a modulational 
instability (also known as the Benjamin-Feir instability \cite{r11,Tracy-Chen} in fluid mechanics) 
of a constant amplitude carrier wave, and indicates that modulationally unstable waves can 
form localized envelope disturbances \cite{r9}. In Refs.\ \cite{Crawford-etal} and \cite{Dysthe-etal}
the effects of a random sea swell on the stability was investigated using the 
single sea state nonlinear deep water wave picture. 

Recently, Onorato {\it et al.} \cite{Onorato-etal} developed a two-dimensional weakly nonlinear model
for two coherent Stokes waves in deep water with two 
different directions of propagation. For one-dimensional wave propagation, Ref. \cite{Onorato-etal} 
presented specific results for the modulational instability of two nonlinearly coupled coherent 
Stokes waves. It was found that freak waves may occur as a result of the modulational instability, 
and the growth rate for the crossing sea state modulational instability was found to be larger 
than that for the single sea state case. Since in reality the waves are incoherent it is however appropriate to examine the effects of random phases using perturbation theory. The presence of a spatial random phase of the background sea state gives rise to spectral broadening of the wave distribution function (as compared to the usual modulational perturbations around a monochromatic background wave distribution), and the modulational instability of nonlinearly coupled Stokes waves can be affected accordingly. This is the objective of the present Letter.
 
In the following, we thus analyze the statistical properties of one-dimensional incoherent crossing 
see states.  In particular, we investigate the modulational instability properties of these 
nonlinearly interacting waves. A generalized distribution function, which is valid for 
partially coherent waves, will be obtained. Using perturbation theory, it will be shown that partial coherence in terms of a random phase approximation gives rise to spectral broadening of 
the background crossing sea states. This broadening tends to stabilize the inherent system modulational instability, enabling 
the interaction between the waves over longer distances before any perturbation grows to appreciable 
levels forming freak waves. In fact, for a wide enough spectral distribution of waves the modulational 
instability can be suppressed. 

In Ref.\ \cite{Onorato-etal} the equations 

\begin{equation}
  \frac{\partial A}{\partial t} - i\alpha\frac{\partial^2A}{\partial x^2} 
  + i(\xi|A|^2 + 2\zeta|B|^2)A = 0  \label{eq:A} 
\end{equation}
and

\begin{equation}
    \frac{\partial B}{\partial t} - i\alpha\frac{\partial^2B}{\partial x^2}  
    + i(\xi|B|^2 + 2\zeta|A|^2)B = 0 , \label{eq:B}
\end{equation}
for one-dimensional comoving crossing sea states were presented.  Here $A$ and $B$ represents the 
surface elevation of the crossing sea states, $\alpha$ is the group velocity dispersion parameter, 
$\xi$ is the nonlinear self-interaction parameter, and $\zeta$ determines the strength of the 
nonlinear interaction between the crossing sea states. We note that the sign of $\alpha$ can always 
be chosen positive by a change of time coordinate. Thus, we take $\alpha > 0$, while the nonlinear 
parameters may be positive as well as negative. It can also be noted that the one-dimensional 
nonlinear Schr\"odinger equation has been shown to be in excellent agreement with laboratory 
experiments \cite{Yuen-Lake}.

There are traveling pure Stokes wave solutions $ A = A_0\exp(-i\omega_A t)$ and $B = B_0\exp(-i\omega_B t)$ of 
Eqs.\ (\ref{eq:A}) and (\ref{eq:B}), where $\omega_A = \xi A_0^2 +2\zeta B_0^2$ and 
$\omega_B = \xi B_0^2 + 2\zeta A_0^2$.  Letting $A(t,x) = A_0[1 + a(t,x)]\exp(-i\omega_At)$ and 
$B(t,x) = B_0 [1 + b(t,x)]\exp(-i\omega_Bt)$, where $a,b \ll 1$, we linearize Eqs.\ (\ref{eq:A}) 
and (\ref{eq:B}), take the real and imaginary parts of the resultant equations, and Fourier analyze 
them against the frequency $\Omega$ and wavenumber $K$ of the modulations. The resulting dispersion 
relation is \cite{Onorato-etal}

\begin{equation}
  \Omega^4 -  
    2\alpha K^2(\alpha K^2 + \xi A_0^2 + \xi B_0^2)\Omega^2  
      + \alpha^2K^4\left[(\alpha K^2 + 2\xi A_0^2)(\alpha K^2 + 2\xi B_0^2) 
       - 16\zeta^2A_0^2B_0^2 \right] = 0 ,
\end{equation}
i.e.\

\begin{equation}\label{eq:disprelmono}
  \Omega = \pm \left\{
     \alpha K^2\left[\alpha K^2 + \xi (A_0^2 + B_0^2) \right] 
     \pm \alpha K^2\sqrt{ 16\zeta^2A_0^2B_0^2 + \xi^2(A_0^2 - B_0^2)^2 } \right\}^{1/2} .
\end{equation}

We now investigate the modulational instability of incoherent crossing sea states by applying 
a Wigner transform to the wave amplitudes $A$ and $B$. For a given function $f(t,x)$, the 
corresponding Wigner function $\rho_f(t,x,p)$ is defined as the Fourier transform of the 
two-point correlation function, i.e.\ \cite{Wigner,Moyal,Alber,Fedele-Anderson,Fedele-etal,Mendonca,Onorato-etal2}

\begin{equation}\label{eq:wignertransf}
  \rho_f(t,x,p) = \frac{1}{2\pi}\int_{-\infty}^{\infty}d\mu\,e^{ip\mu}\langle 
    f^*(t,x + \mu/2)f(t,x-\mu/2) \rangle ,
\end{equation}
where the asterisk stands for the complex conjugate and the angular bracket denotes the ensemble 
average \cite{Klimontovich}. The function $\rho_f$ is a generalized distribution function for 
the waves of the sea states.  The wave intensity $I_f$ corresponding to the function $f$ can 
then be written as

\begin{equation}
  I_f = \langle|f|^2\rangle = \int\,dp\,\rho_f(t,x,p) .
\end{equation}

Applying the time derivative to Eq.\ (\ref{eq:wignertransf}), with $f = A$ or $B$, and using the 
nonlinear Schr\"odinger equations (\ref{eq:A}) and (\ref{eq:B}), valid for slowly varying envelopes, 
respectively, we obtain the nonlinearly coupled wave-kinetic equations

\begin{equation}\label{eq:kineticA}
  \frac{\partial\rho_A}{\partial t} + 2\alpha p\frac{\partial\rho_A}{\partial x} -
  2(\xi I_A + 2\zeta I_B)\sin\left( \frac{1}{2}
    \stackrel{\leftarrow}{\frac{\partial}{\partial x}}%
    \stackrel{\rightarrow}{\frac{\partial}{\partial p}}  \right) \rho_A = 0 ,
\end{equation}
and 

\begin{equation}\label{eq:kineticB}
  \frac{\partial\rho_B}{\partial t} + 2\alpha p\frac{\partial\rho_B}{\partial x} -
  2(\xi I_B + 2\zeta I_A)\sin\left( \frac{1}{2}
    \stackrel{\leftarrow}{\frac{\partial}{\partial x}}%
    \stackrel{\rightarrow}{\frac{\partial}{\partial p}} \right) \rho_B = 0 ,
 \end{equation}
where the $\sin$-operator is defined in terms of its Taylor expansion and the arrows give the direction 
of the differentiation. Equations (\ref{eq:kineticA}) and (\ref{eq:kineticB}) model the nonlinear evolution 
of partially coherent crossing sea states.  

First order perturbations of (\ref{eq:kineticA}) and (\ref{eq:kineticB}) can be treated by letting 
$\rho_j(t,x,p) = \rho_{j0}(p) + \rho_{j1}(p)\exp(iKx - i\Omega t)$ and $I_j(t,x) 
= I_{j0} + I_{j1}\exp(iKx - i\Omega t)$, where $j = A, B$, $|\rho_{j1}| \ll \rho_{j0}$, 
and $|I_{j1}| \ll I_{j0}$. Linearizing with respect to the perturbation variables, we obtain 
the dispersion relation from Eqs.\ (\ref{eq:kineticA}) and (\ref{eq:kineticB})

\begin{equation}\label{eq:disprelkinetic}
  (1 + \xi\Delta_A)(1 + \xi\Delta_B) - 4\zeta^2\Delta_A\Delta_B = 0, 
\end{equation}
where 

\begin{equation}\label{eq:delta}
  \Delta_j = \int\,dp\,\frac{\rho_{j0}(p + K/2) - \rho_{j0}(p - K/2)}{\Omega - 2\alpha Kp} .
\end{equation}
Here we have used $2i\sin(iK/2\partial_p)\rho_{j0}(p) = -\rho_{j0}(p + K/2) + \rho_{j0}(p - K/2)$.
The dispersion relation (\ref{eq:disprelkinetic}) with (\ref{eq:delta}) is valid for partially coherent 
crossing sea states that interact nonlinearly. 

In the coherent case, the background waves are simple plane waves, so that the distribution functions are given by $\rho_{j0} = I_{j0}\delta(p)$, 
where $I_{A0} = A_0^2$ and $I_{B0} = B_0^2$. Inserting the coherent  background distribution function 
into (\ref{eq:delta}), the dispersion relation (\ref{eq:disprelkinetic}) reduces to Eq.\ (\ref{eq:disprelmono}).

We now assume that the sea states suffer from random perturbations, such that they have partial 
phase coherence. Thus means that the phase $\varphi_{j}(x)$ of the background wave amplitudes $A, B$ satisfies $\langle \exp[-i\varphi_j(x+\mu/2)\exp[i\varphi_j(x-\mu/2)]]\rangle = \exp(-p_{jW}|\mu|)$, corresponding to the Lorentz distribution function \cite{Loudon}

\begin{equation}
  \rho_{j0}(p) = \frac{I_{j0}}{\pi}\frac{p_{jW}}{p^2 + p_{jW}^2},
\end{equation}
where $p_{jW}$ represents the width of the $j$th distribution function. Thus, we see that the partial coherence in the waves phases give rises to a spectral broadening in terms of the distribution function. With this spectral background 
of sea states, the dispersion relation (\ref{eq:disprelkinetic}) takes the form

\begin{eqnarray}
  && \left[ 1 - \frac{2\xi\alpha K^2 I_{A0}}{(\Omega + 2i\alpha Kp_{AW})^2 - \alpha^2K^4} \right]%
  \left[ 1 - \frac{2\xi\alpha K^2 I_{B0}}{(\Omega + 2i\alpha Kp_{BW})^2 - \alpha^2K^4} \right]
  \nonumber \\ &&\qquad\qquad
  - \frac{16\zeta^2\alpha^2 K^4 I_{A0}I_{B0}}{[(\Omega + 2i\alpha Kp_{AW})^2 - \alpha^2K^4]%
  [(\Omega + 2i\alpha Kp_{BW})^2 - \alpha^2K^4]} = 0 .
\label{eq:disprellorentz}
\end{eqnarray} 
We next analyze the dispersion relation (\ref{eq:disprellorentz}) by letting, for simplicity, 
$p_{AW} = p_{BW} = p_W$.  We then obtain 

\begin{equation}\label{eq:disprelfinal}
   \Omega = -2i\alpha K p_W \pm \left\{
     \alpha K^2\left[\alpha K^2 + \xi (I_{A0} + I_{B0}) \right] 
     \pm \alpha K^2\sqrt{ 
       16\zeta^2I_{A0}I_{B0} + \xi^2(I_{A0} - I_{B0})^2 } \right\}^{1/2} ,
\end{equation}
which agrees with (\ref{eq:disprelmono}) when $p_W \rightarrow 0$. We may then transform to dimensionless 
variables by $\Omega \rightarrow \Omega/\sqrt{\alpha|\xi|}$, $\alpha K^2 \rightarrow \alpha K^2/\sqrt{\alpha|\xi|}$, 
$\alpha K p_W \rightarrow \alpha K p_W/\sqrt{\alpha|\xi|}$, $I_{j0} \rightarrow I_{j0}\sqrt{|\xi|/\alpha}$, 
and $\zeta \rightarrow \zeta/\xi$. With these re-scalings, the group velocity dispersion  
$\alpha = 1$, while the self-nonlinearity coefficient becomes $\xi \rightarrow \pm 1$. In Figs.\ 1 and 2 we have 
displayed the dimensionless growth rate $\Gamma = -i\Omega$ using Eq.\ (\ref{eq:disprelfinal}). In Fig.\ 1
we have chosen $I_{A0} = I_{B0} = 0.5$, while in Fig.\ 2 we have $I_{A0} = 5I_{B0} = 0.5$. The lowering of 
the growth rate due to the spectral broadening of the background sea states $\rho_{j0}$ can clearly be seen. 
Thus, the effect of the partial coherence of the background sea states is to stabilize their dynamics, 
allowing for long distance propagation of envelope solitons without the growth of freak waves. It should be 
noted that in general the spectral broadening enables interaction between the ocean waves over longer 
distances before any initial perturbation grows to an appreciable size. However, for large enough spectral 
width $p_W$ the modulational instability growth rate may be completely suppressed. The results for 
$p_W = 0$, i.e.\ for coherent background waves without spatial spectral broadening, are consistent with the results 
presented by Onorato, Osborne, and Serio \cite{Onorato-etal} (see the full curves in Figs.\ 1 and 2), 
where it was found that crossing sea states have a larger growth rate for the instability as compared 
to the single sea state instability.    

%%%%%%%%% FIG %%%%%%%%%
\begin{figure}
\subfigure[]{\includegraphics[width=0.48\textwidth]{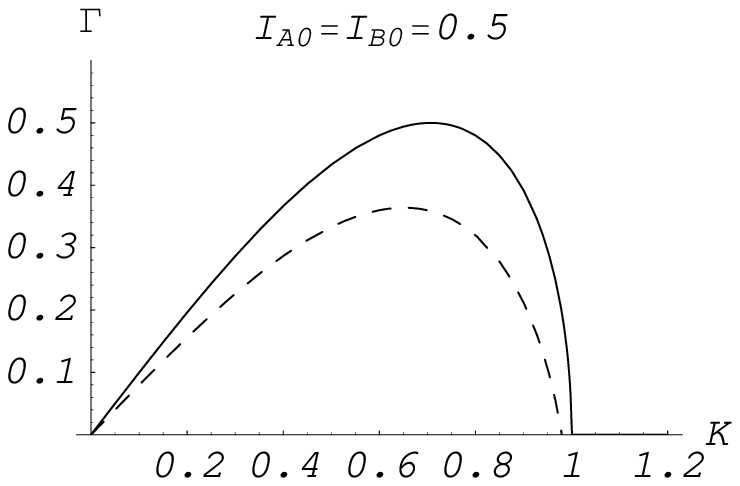}}
\subfigure[]{\includegraphics[width=0.48\textwidth]{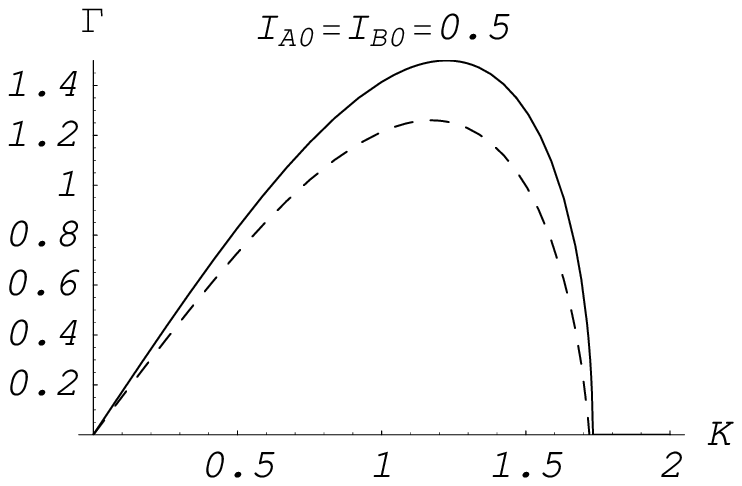}}
\subfigure[]{\includegraphics[width=0.48\textwidth]{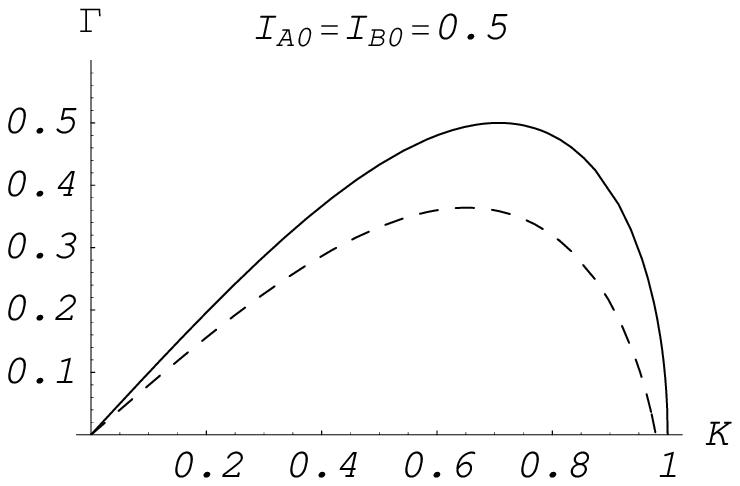}}
\subfigure[]{\includegraphics[width=0.48\textwidth]{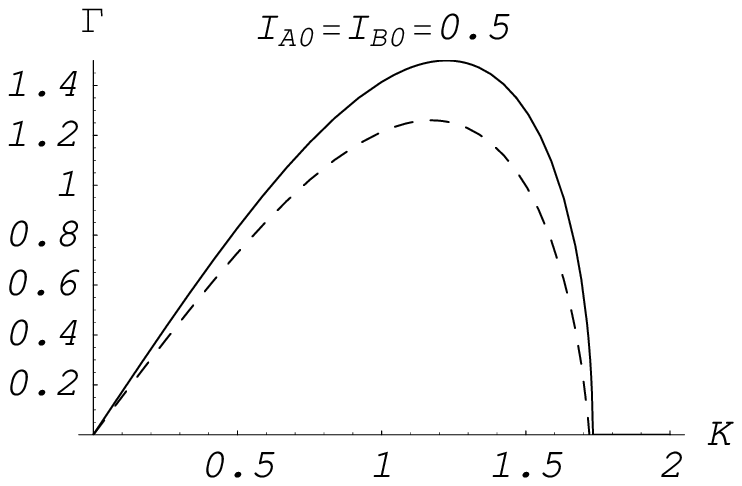}}
\caption{The effect of partial coherence when $\alpha$ is normalized to unity and $I_{A0} = I_{B0} = 0.5$. 
In all the panels, the coherent (full) and incoherent (dashed) 
cases are compared. In (a) $\zeta = \xi =1$, in (b) $\zeta = -\xi = 1$, in (c) $\zeta = -\xi = -1$, 
and in (d) $\zeta = \xi = -1$.  
The partial coherence stabilizes the wave modulation, i.e.\ the wave will be able to interact
over longer distances before the perturbations grow to appreciable levels forming freak waves. 
In fact, for a wide enough spectral distribution, the modulational instability can 
be \emph{completely suppressed.}}
\end{figure}
%%%%%%%%%%%%%%%%%%%%%

%%%%%%%%% FIG %%%%%%%%%
\begin{figure}
\subfigure[]{\includegraphics[width=0.48\textwidth]{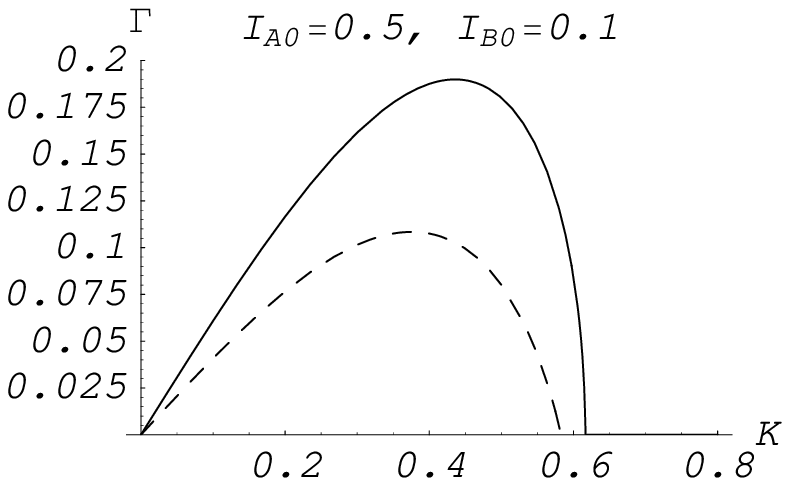}}
\subfigure[]{\includegraphics[width=0.48\textwidth]{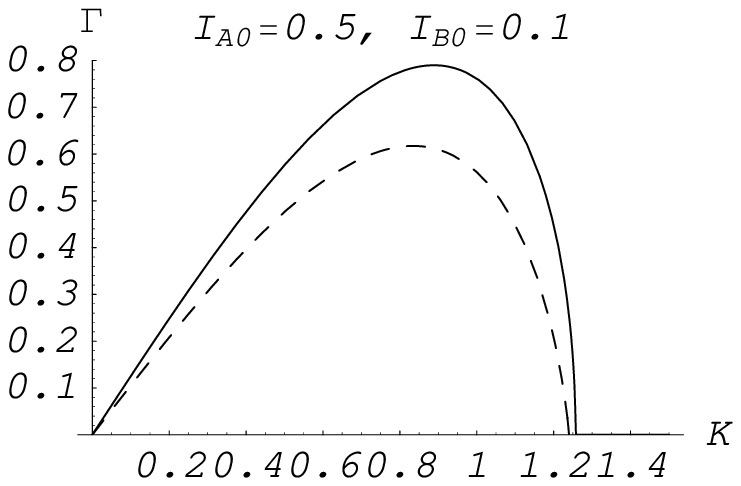}}
\subfigure[]{\includegraphics[width=0.48\textwidth]{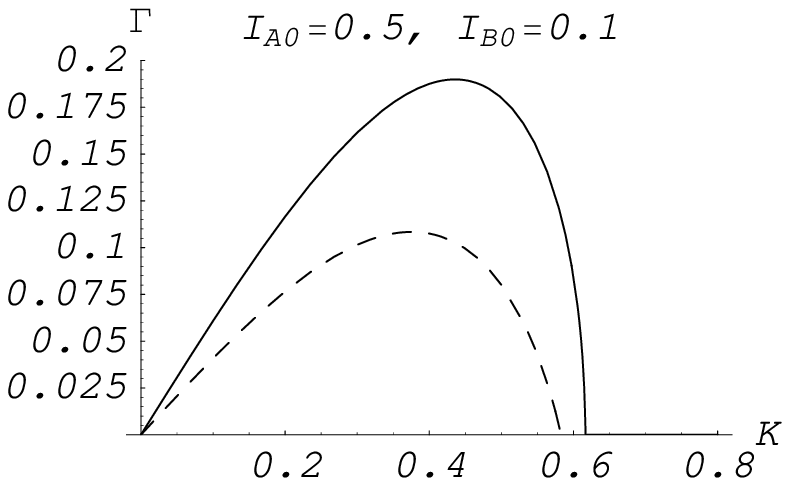}}
\subfigure[]{\includegraphics[width=0.48\textwidth]{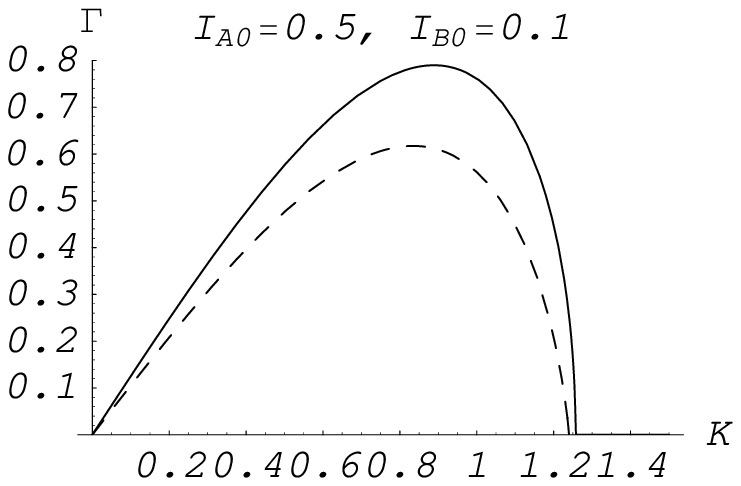}}
\caption{The effect of partial coherence when $\alpha$ is normalized to unity and $I_{A0} = 5I_{B0} = 0.5$. 
In all the panels, the coherent (full) and incoherent (dashed) 
cases are compared.  In (a) $\zeta = \xi =1$, in (b) $\zeta = -\xi = 1$, in (c) $\zeta = -\xi = -1$, 
and in (d) $\zeta = \xi = -1$.  
As in the cases in Fig.\ 1, the partial coherence stabilizes the sea state modulation.}
\end{figure}
%%%%%%%%%%%%%%%%%%%%%

Indeed, there are soliton solutions to the system (\ref{eq:A}) and (\ref{eq:B}). 
Following Ref.\ \cite{Vladimirov-etal}, we can find the dark--bright soliton pair
\begin{equation}
  A(t,x) = A_0\tanh[X(t,x)]\exp(i\kappa x - i\delta\, t) 
\end{equation}
and 
\begin{equation}
  B(t,x) = B_0\,\mathrm{sech}[X(t,x)]\exp(i\kappa x - i\delta\,t)  ,
\end{equation}
where $X = (x - Vt)/L$, $V = 2\alpha\kappa$, $A_0^2 = B_0^2$, and $\delta = \alpha\kappa^2 + \xi A_0^2$. 
Here $A_0$ and $L$ are treated as two free parameters determining the elevation and width of the solitons, 
respectively. Of course, the classical nonlinear Schr\"odinger equation has both soliton and multi-soliton solutions. Here we see that also the co-existence of dark and bright soliton water waves is possible. 

To summarize, we have presented an investigation of the modulational instability of two incoherent
crossing sea states that are nonlinearly interacting in deep water. For this purpose, we have introduced the 
Wigner transformation of the coupled nonlinear Schr\"odinger equations of Ref. \cite{Onorato-etal}
and obtained two coupled wave-kinetic equations (or von Neumann equations). The latter have been  
analyzed to obtain a nonlinear dispersion relation for background sea states that have broadband spectra, i.e.  finite spectral width.
It is found that the growth rate of the modulational instability is suppressed.
Hence, random phased nonlinearly interacting waves could propagate over long 
distances without being much affected by the modulational instability, and for a wide enough spectral 
distribution the formation of freak waves is thus completely suppressed. 
However, it should be stressed that this complete suppression is a result of the above NLS model calculations. 
As an alternative, one could start from the original equations,
and then consider the interaction of wave packets which are eigenfunctions of the 
corresponding linearised problem with a background of Stokes wave trains.
In that case, a random mixture of the eigenfunctions will postpone, although 
not completely suppress, the freak wave appearance.

\acknowledgments
This research was partially supported by the Swedish Research Council.

\end{document}